\documentclass[pdflatex,sn-mathphys-num]{sn-jnl}

\pdfmapfile{=cm.map}
\pdfmapfile{+symbols.map}
\pdfmapfile{+cmextra.map}

\usepackage{graphicx}
\usepackage{amsmath,amssymb,amsfonts}
\usepackage{amsthm}
\usepackage{mathtools}
\usepackage{bm}
\usepackage{booktabs}

\newcommand{\ket}[1]{\lvert #1\rangle}
\newcommand{\bra}[1]{\langle #1\rvert}
\newcommand{\proj}[1]{\lvert #1\rangle\!\langle #1\rvert}
\newcommand{\Tr}{\operatorname{tr}}
\newcommand{\spec}{\operatorname{spec}}
\newcommand{\id}{\operatorname{id}}

\theoremstyle{thmstyleone}
\newtheorem{theorem}{Theorem}
\newtheorem{proposition}[theorem]{Proposition}
\newtheorem{corollary}[theorem]{Corollary}

\theoremstyle{thmstyletwo}
\newtheorem{example}{Example}
\newtheorem{remark}{Remark}

\theoremstyle{thmstylethree}
\newtheorem{definition}{Definition}

\begin{document}

\title[Parity-reflected complementarity]{Parity-Reflected Complementarity and Quantum-Capacity Regimes of Hadamard-Convolution Channels}

\author[1]{\fnm{Laizhen} \sur{Luo}}

\author*[2]{\fnm{Chunhe} \sur{Xiong}}\email{xiongchunhe@csu.edu.cn}

\affil[1]{\orgdiv{Foundation Science Education Center},
  \orgname{Hangzhou City University},
  \orgaddress{\city{Hangzhou}, \postcode{310015}, \country{China}}}

\affil*[2]{\orgdiv{School of Mathematics and Statistics},
  \orgname{Central South University},
  \orgaddress{\city{Changsha}, \postcode{410083}, \country{China}}}

\abstract{We study the quantum and private capacities of finite-dimensional Hadamard-convolution channels generated by coupling an input to a fixed environmental state. First, we prove that the weak complementary channel is another Hadamard-convolution channel whose environment is obtained by parity reflection. Hence, for pure environments, the same statement holds for the ordinary complementary channel. This identity also yields explicit antidegradability criteria and establishes vanishing quantum and private capacities for broad symmetry classes, including an asymmetric nonstabilizer environment. Independently, we construct a qutrit family of positive-partial-transpose Choi states, and therefore have zero quantum capacity. Conversely, we give an asymmetric nonstabilizer qutrit environment with strictly positive coherent information, proving positive quantum capacity and excluding universal antidegradability. The coexistence of magic zero-capacity and positive-capacity examples separates state-level resources from coupling-induced channel structure. These results show that environmental magic alone does not determine communication performance. Instead, parity-reflected complementarity provides a channel-level mechanism separating zero- and positive-capacity regimes.}

\keywords{quantum convolution, quantum capacity, antidegradable channel, coherent information, magic states, positive-partial-transpose channel}

\maketitle

\section{Introduction}\label{sec:introduction}

The quantum capacity of a noisy channel quantifies the asymptotic rate at which quantum information can be transmitted with vanishing error. Its regularized nature makes exact evaluation difficult: coherent information can be superadditive \cite{DiVincenzo1998pra,Leditzky2018prl}, and even channels with vanishing individual quantum capacities can exhibit superactivation \cite{Smith2008science}. Structural channel properties therefore play a central role in capacity theory. Antidegradable channels have zero quantum and private capacities because the receiver's output can be simulated from the environment. Channels whose Choi states have positive partial transpose (PPT) provide another important class with zero quantum capacity \cite{Smith2012prl,Horodecki1998prl}.

Discrete quantum convolution provides a structured family of system--environment interactions in which a channel is generated by coupling an input state to a fixed environmental state through a finite-field linear transformation. The framework was introduced by Bu, Gu, and Jaffe in their study of quantum entropy, discrete quantum Gaussians, and quantum central limit theorems \cite{Bu2023pnas,Bu2023arxiv}. In this setting, stabilizer states play a role analogous to that of Gaussian states in continuous-variable systems, while nonstabilizer states provide the corresponding magic resource.

For discrete beam-splitter convolutions, Bu and Jaffe showed that stabilizer environments lead to zero quantum capacity, whereas suitable magic environments can yield positive quantum capacity; they also bounded the capacity in terms of environmental magic \cite{Bu2025prl}. Sun and Jin subsequently extended the capacity analysis to broader classes of stabilizer convolutions and to lower-dimensional systems \cite{Sun2025pra}. Related work has examined the private capacity of convolutional channels and shown that environmental magic can support private communication, while certain symmetric magic environments still produce zero private capacity \cite{Xiong2026arxiv}.

These results motivate a structural question that is different from merely asking whether an environmental state is stabilizer or magic: how is the complementary channel determined by the environment and by the coupling? We answer this question for the Hadamard-convolution, whose two outputs are the finite-field sum and difference of the two inputs. The answer is an exact parity-reflection identity: the weak complementary output is generated by replacing the environment with its parity-reflected state. This identity makes it possible to formulate antidegradability directly in terms of environmental reflection symmetry.

The main contributions are as follows. First, we prove that the weak complementary channel generated by an environment is the Hadamard-convolution channel generated by its parity-reflected state. For pure environments, the equality holds for the ordinary complementary channel. Second, we show that parity-invariant environments, and more generally channel-reflection-equivalent environments, generate antidegradable channels. Their quantum and private capacities therefore vanish. An explicit asymmetric nonstabilizer example demonstrates that magic alone is not sufficient for positive communication capacity. Third, we construct a family of one-qutrit environments for which the normalized Choi state is PPT. The corresponding channels have zero quantum capacity, although the PPT condition alone does not imply zero private capacity. Finally, we exhibit an asymmetric nonstabilizer qutrit environment and an input state with strictly positive coherent information. Consequently, the corresponding channel has positive quantum capacity, proving that Hadamard-convolution channels are not universally antidegradable.

The remainder of the paper is organized as follows. Section~\ref{sec:framework} introduces Hadamard-convolution channels and the relevant capacity quantities. Section~\ref{sec:reflection} proves the parity-reflected complementarity identity. Section~\ref{sec:zero} develops the antidegradability and PPT zero-capacity mechanisms. Section~\ref{sec:positive} establishes positive coherent information for an explicit qutrit channel. Section~\ref{sec:conclusion} summarizes the results and discusses the remaining classification problem.

\section{Hadamard-convolution channels}\label{sec:framework}

\subsection{Qudit notation and channel capacities}\label{subsec:notation}

All logarithms are taken to base two. Let $d$ be an odd prime and let $\mathcal H=\mathbb C^d$. We denote the computational basis by $\{\ket{k}:k\in\mathbb Z_d\}$ and define the generalized Pauli operators by
\begin{equation*}
 X\ket{k}=\ket{k+1},
 \qquad
 Z\ket{k}=\omega_d^k\ket{k},
 \qquad
 \omega_d=e^{2\pi i/d}.
\end{equation*}
For $n$ qudits, we write $\ket{\bm k}=\ket{k_1}\otimes\cdots\otimes\ket{k_n}$, and all vector arithmetic is performed componentwise modulo $d$. The parity-reflection unitary is
\begin{equation}
 \Pi\ket{\bm k}=\ket{-\bm k}.
\label{eq:parity}
\end{equation}

Let $V:\mathcal H_A\to\mathcal H_B\otimes\mathcal H_E$ be an isometric extension of a channel $\Lambda$. The channel and a complementary channel are
\begin{equation*}
 \Lambda(\rho)=\Tr_E(V\rho V^\dagger),
 \qquad
 \Lambda^c(\rho)=\Tr_B(V\rho V^\dagger).
\end{equation*}
For an input ensemble $\mathcal E=\{p_x,\rho_x\}$, its Holevo information at the output of $\Lambda$ is
\begin{equation*}
 \chi(\mathcal E;\Lambda)
 =S\!\left(\sum_xp_x\Lambda(\rho_x)\right)
  -\sum_xp_xS(\Lambda(\rho_x)).
\end{equation*}
The one-shot private information and optimized coherent information are
\begin{align}
 P^{(1)}(\Lambda)
  &=\max_{\mathcal E}
    \bigl[\chi(\mathcal E;\Lambda)-\chi(\mathcal E;\Lambda^c)\bigr],
 \label{eq:private-one-shot}\\
 Q^{(1)}(\Lambda)
  &=\max_\rho
    \bigl[S(\Lambda(\rho))-S(\Lambda^c(\rho))\bigr].
 \label{eq:coherent-one-shot}
\end{align}
The private and quantum capacities are obtained by regularization \cite{Lloyd1997pra,Shor2002msri,Devetak2005tit}:
\begin{equation}
 P(\Lambda)=\sup_{k\geq1}\frac{1}{k}P^{(1)}(\Lambda^{\otimes k}),
 \qquad
 Q(\Lambda)=\sup_{k\geq1}\frac{1}{k}Q^{(1)}(\Lambda^{\otimes k}).
\label{eq:regularized-capacities}
\end{equation}
A channel is degradable if $\Lambda^c=\mathcal D\circ\Lambda$ for some channel $\mathcal D$, and antidegradable if $\Lambda=\mathcal A\circ\Lambda^c$ for some channel $\mathcal A$ \cite{Cubitt2008jmp}. Tensor powers of an antidegradable channel remain antidegradable. Data processing then gives $Q=P=0$.

\subsection{Hadamard convolution and its weak complement}\label{subsec:channel}

The $n$-qudit Hadamard-convolution unitary $U_H:\mathcal H_A^{\otimes n}\otimes\mathcal H_E^{\otimes n}\to\mathcal H_B^{\otimes n}\otimes\mathcal H_F^{\otimes n}$ is
\begin{equation}
 U_H
 =\sum_{\bm i,\bm j\in\mathbb Z_d^n}
   \ket{\bm i,\bm j}\bra{\bm i+\bm j,\bm i-\bm j}.
\label{eq:UH}
\end{equation}
Equivalently,
\begin{equation}
 U_H\ket{\bm x,\bm y}
 =\ket{2^{-1}(\bm x+\bm y),,2^{-1}(\bm x-\bm y)},
\label{eq:UH-action}
\end{equation}
where $2^{-1}$ is the multiplicative inverse of $2$ in $\mathbb Z_d$. The assumption that $d$ is odd ensures that the underlying linear transformation is invertible and hence that $U_H$ is unitary.

Given an environmental state $\sigma$ on the second input, the Hadamard-convolution channel is
\begin{equation}
 \Lambda_\sigma(\rho)
 =\Tr_F\!\left[U_H(\rho\otimes\sigma)U_H^\dagger\right].
\label{eq:main-channel}
\end{equation}
We also define its weak complementary channel by retaining the other output of the same coupling:
\begin{equation}
 \widetilde{\Lambda}^{\,c}_\sigma(\rho)
 =\Tr_B\!\left[U_H(\rho\otimes\sigma)U_H^\dagger\right].
\label{eq:weak-complement}
\end{equation}
If $\sigma$ is pure, Eq.~\eqref{eq:weak-complement} is an ordinary complementary channel. If $\sigma$ is mixed, let $\ket{\Psi_\sigma}_{ER}$ be a purification of $\sigma_E$. The ordinary complementary channel has output $FR$, and tracing out $R$ yields the weak complement:
\begin{equation}
 \widetilde{\Lambda}^{\,c}_\sigma
 =\Tr_R\circ\Lambda_\sigma^c.
\label{eq:weak-true-relation}
\end{equation}

\begin{remark}[Terminology]\label{rem:terminology}
Throughout this work, a \emph{Hadamard-convolution channel} means the channel induced by the finite-field sum--difference unitary in Eq.~\eqref{eq:UH}. It should not be confused with a \emph{quantum Hadamard channel} in the standard quantum-Shannon-theoretic terminology, whose complementary channel is entanglement breaking \cite{Bradler2010pra,Wilde2014cmp}.
\end{remark}

\section{Parity-reflected complementarity}\label{sec:reflection}

We now establish the central structural identity. The exact equality for pure environments refers to the canonical Stinespring isometry obtained directly from $U_H$ and the environmental vector. As usual, any other complementary channel is related to this one by an isometry on its output.

\begin{theorem}[Parity-reflected complementarity]\label{thm:reflection}
Let $\sigma$ be an $n$-qudit environmental state. Then
\begin{equation}
 \widetilde{\Lambda}^{\,c}_\sigma
 =\Lambda_{\Pi\sigma\Pi^\dagger}.
\label{eq:weak-reflection}
\end{equation}
In particular, if $\sigma=\proj{\psi}$ is pure and $\ket{\psi^\Pi}:=\Pi\ket{\psi}$, then the canonical complementary channel satisfies
\begin{equation}
 \Lambda_\psi^c=\Lambda_{\psi^\Pi}.
\label{eq:pure-reflection}
\end{equation}
\end{theorem}

\begin{proof}
We first consider a pure environmental state
\begin{equation*}
 \ket{\psi}=\sum_{\bm r\in\mathbb Z_d^n}c_{\bm r}\ket{\bm r}
\end{equation*}
and the canonical Stinespring isometry $V_\psi:=U_H(I\otimes\ket{\psi})$. Projecting the second output onto $\ket{\bm m}$ gives Kraus operators for the direct channel:
\begin{align}
 K_{\bm m}^{(\psi)}=(I\otimes\bra{\bm m})V_\psi=\sum_{\bm k\in\mathbb Z_d^n}
   c_{\bm k-\bm m}\ket{\bm k}\bra{\bm k+\bm m}.
\label{eq:direct-kraus}
\end{align}
Thus $\Lambda_\psi(\rho)=\sum_{\bm m}K_{\bm m}^{(\psi)}\rho K_{\bm m}^{(\psi)\dagger}$. Projecting the first output instead gives Kraus operators for the canonical complementary channel:
\begin{align}
 L_{\bm m}^{(\psi)}=(\bra{\bm m}\otimes I)V_\psi=\sum_{\bm k\in\mathbb Z_d^n}c_{\bm m-\bm k}\ket{\bm k}\bra{\bm k+\bm m}.
\label{eq:complement-kraus}
\end{align}
Consequently, $\Lambda_\psi^c(\rho)=\sum_{\bm m}L_{\bm m}^{(\psi)}\rho L_{\bm m}^{(\psi)\dagger}$.

The amplitudes of the parity-reflected state satisfy
\begin{equation*}
 c_{\bm r}^{\Pi}:=\langle\bm r\vert\psi^\Pi\rangle=c_{-\bm r}.
\end{equation*}
Using Eq.~\eqref{eq:direct-kraus}, the Kraus operators generated by $\ket{\psi^\Pi}$ are
\begin{align*}
 K_{\bm m}^{(\psi^\Pi)}=\sum_{\bm k}c_{\bm k-\bm m}^{\Pi}\ket{\bm k}\bra{\bm k+\bm m}=\sum_{\bm k}c_{\bm m-\bm k}
   \ket{\bm k}\bra{\bm k+\bm m}=L_{\bm m}^{(\psi)}.
\end{align*}
Therefore $\Lambda_\psi^c=\Lambda_{\psi^\Pi}$.

For a mixed environment, choose any pure-state decomposition $\sigma=\sum_a p_a\proj{\psi_a}$. Both sides of Eq.~\eqref{eq:weak-reflection} depend linearly on the environmental density operator. Hence
\begin{align*}
 \widetilde{\Lambda}^{\,c}_\sigma=\sum_a p_a\Lambda_{\psi_a}^c
 =\sum_a p_a\Lambda_{\psi_a^\Pi}=\Lambda_{\Pi\sigma\Pi^\dagger},
\end{align*}
which proves the mixed-state identity.
\end{proof}

\section{Zero-capacity mechanisms}\label{sec:zero}

\subsection{Antidegradability from reflection symmetry}\label{subsec:antidegradable}

Theorem~\ref{thm:reflection} converts environmental parity symmetry into a direct antidegradability criterion.

\begin{corollary}[Parity-invariant environments]\label{cor:parity-invariant}
If $\Pi\sigma\Pi^\dagger=\sigma$, then $\Lambda_\sigma$ is antidegradable and
\begin{equation*}
 Q(\Lambda_\sigma)=P(\Lambda_\sigma)=0.
\end{equation*}
If, in addition, $\sigma=\proj{\psi}$ is pure, then $\Lambda_\psi$ is self-complementary and is both degradable and antidegradable.
\end{corollary}

\begin{proof}
Let $\ket{\Psi_\sigma}_{ER}$ be a purification of $\sigma_E$. By Eq.~\eqref{eq:weak-true-relation}, Theorem~\ref{thm:reflection}, and parity invariance,
\begin{equation*}
 \Tr_R\circ\Lambda_\sigma^c
 =\widetilde{\Lambda}^{\,c}_\sigma
 =\Lambda_{\Pi\sigma\Pi^\dagger}
 =\Lambda_\sigma.
\end{equation*}
Thus the channel $\mathcal A=\Tr_R$ maps the ordinary complementary output to the receiver's output, which proves antidegradability. For every input ensemble, data processing gives
\begin{equation*}
 \chi(\mathcal E;\Lambda_\sigma)
 \leq \chi(\mathcal E;\Lambda_\sigma^c).
\end{equation*}
The same relation holds for all tensor powers, so Eqs.~\eqref{eq:private-one-shot} and \eqref{eq:regularized-capacities} give $P(\Lambda_\sigma)=0$; antidegradability likewise gives $Q(\Lambda_\sigma)=0$. If $\sigma$ is pure, the purification system $R$ is absent and $\Lambda_\psi^c=\Lambda_\psi$, so the channel is self-complementary.
\end{proof}

Parity invariance is sufficient but not necessary. The next definition captures a larger class.

\begin{definition}[Channel-reflection equivalence]\label{def:reflection-equivalence}
An environmental state $\sigma$ is \emph{channel-reflection equivalent} if there exists a unitary channel $\mathcal W(\tau)=W\tau W^\dagger$ such that
\begin{equation}
 \Lambda_{\Pi\sigma\Pi^\dagger}=\mathcal W\circ\Lambda_\sigma.
\label{eq:reflection-equivalence}
\end{equation}
\end{definition}

\begin{proposition}\label{prop:reflection-antidegradable}
If $\sigma$ is channel-reflection equivalent, then $\Lambda_\sigma$ is antidegradable and $Q(\Lambda_\sigma)=P(\Lambda_\sigma)=0$. If $\sigma$ is pure, then $\Lambda_\sigma$ is both degradable and antidegradable.
\end{proposition}

\begin{proof}
Combining Eqs.~\eqref{eq:weak-true-relation}, \eqref{eq:weak-reflection}, and \eqref{eq:reflection-equivalence} yields
\begin{equation*}
 \Tr_R\circ\Lambda_\sigma^c
 =\mathcal W\circ\Lambda_\sigma.
\end{equation*}
Therefore
\begin{equation*}
 \Lambda_\sigma
 =\mathcal W^\dagger\circ\Tr_R\circ\Lambda_\sigma^c.
\end{equation*}
The map $\mathcal A:=\mathcal W^\dagger\circ\Tr_R$ is completely positive and trace preserving, so $\Lambda_\sigma$ is antidegradable. If $\sigma$ is pure, $R$ is absent and $\Lambda_\sigma^c=\mathcal W\circ\Lambda_\sigma$. The unitary channels $\mathcal W$ and $\mathcal W^\dagger$ then provide degrading and antidegrading maps, respectively.
\end{proof}

\begin{example}[An asymmetric nonstabilizer zero-capacity environment]\label{ex:magic-zero}
Let $d$ be an odd prime and consider a single qudit
\begin{equation*}
 \ket{\psi_+}=\frac{\ket{0}+\ket{1}}{\sqrt2}.
\end{equation*}
A pure single-qudit stabilizer state in odd prime dimension has either one nonzero computational-basis amplitude or $d$ amplitudes of equal modulus \cite{Gross2006jmp}. Hence $\ket{\psi_+}$, which has exactly two nonzero amplitudes, is nonstabilizer. It is also not parity invariant.

For every $\bm a\in\mathbb Z_d^n$, Eq.~\eqref{eq:UH-action} gives the environment-translation covariance relation
\begin{equation}
 U_H(I\otimes X_{\bm a})
 =\left(X_{2^{-1}\bm a}\otimes X_{-2^{-1}\bm a}\right)U_H.
\label{eq:unitary-translation}
\end{equation}
After tracing out the second output,
\begin{equation}
 \Lambda_{X_{\bm a}\sigma X_{\bm a}^\dagger}=\mathcal X_{2^{-1}\bm a}\circ\Lambda_\sigma,
\label{eq:channel-translation}
\end{equation}
where $\mathcal X_{\bm b}(\tau):=X_{\bm b}\tau X_{\bm b}^\dagger.$ Since $\Pi\ket{\psi_+}=X_{-1}\ket{\psi_+}$, Eq.~\eqref{eq:channel-translation} implies
\begin{equation*}
 \Lambda_{\Pi\psi_+\Pi^\dagger}
 =\mathcal X_{-2^{-1}}\circ\Lambda_{\psi_+}.
\end{equation*}
Thus $\ket{\psi_+}$ is channel-reflection equivalent. Proposition~\ref{prop:reflection-antidegradable} gives
\begin{equation*}
 Q(\Lambda_{\psi_+})=P(\Lambda_{\psi_+})=0.
\end{equation*}
This example shows that environmental magic alone does not ensure positive quantum or private capacity.
\end{example}

\subsection{A qutrit family of PPT channels}\label{subsec:ppt}

Antidegradability is not the only zero-quantum-capacity mechanism in this channel family. We next construct a qutrit family whose normalized Choi state is PPT. Let
\begin{equation}
 \ket{\Phi_3}=\frac{1}{\sqrt3}\sum_{j=0}^2\ket{j,j},
 \qquad
 J(\Lambda_\sigma)
 =(\id\otimes\Lambda_\sigma)(\proj{\Phi_3}).
\label{eq:normalized-choi}
\end{equation}
Write $\sigma=\sum_{r,s=0}^2\sigma_{rs}\ket r\bra s$. From Eq.~\eqref{eq:UH-action}, for $x,u,\Delta\in\mathbb Z_3$ one obtains
\begin{equation}
 \Lambda_\sigma(\ket x\bra{x+\Delta})
 =\sum_{u\in\mathbb Z_3}
 \sigma_{2u-x,\,2u-x+\Delta}
 \ket u\bra{u+\Delta}.
\label{eq:qutrit-matrix-action}
\end{equation}
Consequently, partial transposition on the channel-output factor gives
\begin{equation}
 J(\Lambda_\sigma)^{T_B}
 =\frac13\sum_{x,u,\Delta\in\mathbb Z_3}
 \sigma_{2u-x,\,2u-x+\Delta}
 \ket{x,u+\Delta}\bra{x+\Delta,u}.
\label{eq:choi-partial-transpose-sum}
\end{equation}
The sum of the two computational-basis indices is the same in the ket and bra of every term in Eq.~\eqref{eq:choi-partial-transpose-sum}. Hence the operator preserves the three subspaces of fixed index sum modulo $3$. In the ordered basis
\begin{equation*}
 \bigl(\ket{00},\ket{12},\ket{21};\
       \ket{02},\ket{11},\ket{20};\
       \ket{01},\ket{10},\ket{22}\bigr),
\end{equation*}
it therefore has the block decomposition
\begin{equation}
 P J(\Lambda_\sigma)^{T_B}P^\dagger
 =\frac13\left(C_0\oplus C_1\oplus C_2\right),
\label{eq:choi-blocks}
\end{equation}
where $P$ is the corresponding permutation unitary and
\begin{align}
 C_0&=
 \begin{pmatrix}
 \sigma_{00}&\sigma_{12}&\sigma_{21}\\
 \sigma_{21}&\sigma_{00}&\sigma_{12}\\
 \sigma_{12}&\sigma_{21}&\sigma_{00}
 \end{pmatrix},
 &
 C_1&=
 \begin{pmatrix}
 \sigma_{11}&\sigma_{20}&\sigma_{02}\\
 \sigma_{02}&\sigma_{11}&\sigma_{20}\\
 \sigma_{20}&\sigma_{02}&\sigma_{11}
 \end{pmatrix},
 \label{eq:C0C1}\\[1ex]
 C_2&=
 \begin{pmatrix}
 \sigma_{22}&\sigma_{01}&\sigma_{10}\\
 \sigma_{10}&\sigma_{22}&\sigma_{01}\\
 \sigma_{01}&\sigma_{10}&\sigma_{22}
 \end{pmatrix}.
 \label{eq:C2}
\end{align}
Each block is Hermitian and circulant. This decomposition reduces the PPT test to three $3\times3$ matrices.

\begin{proposition}[A qutrit PPT family]\label{prop:ppt-family}
Let $d=3$, $n=1$, and
\begin{equation}
 \sigma=
 \begin{pmatrix}
 a&0&\sqrt{ac}\\
 0&b&0\\
 \sqrt{ac}&0&c
 \end{pmatrix},
 \qquad
 a,b,c>0,
 \qquad
 a+b+c=1.
\label{eq:ppt-environment}
\end{equation}
If $b\geq\sqrt{ac}$, then $J(\Lambda_\sigma)^{T_B}\geq0$. Consequently,
\begin{equation*}
 Q(\Lambda_\sigma)=0.
\end{equation*}
\end{proposition}

\begin{proof}
The $0$--$2$ principal block of $\sigma$ is
\begin{equation*}
 \begin{pmatrix}a&\sqrt{ac}\\\sqrt{ac}&c\end{pmatrix}
 =\begin{pmatrix}\sqrt a\\\sqrt c\end{pmatrix}
  \begin{pmatrix}\sqrt a&\sqrt c\end{pmatrix},
\end{equation*}
so $\sigma\geq0$, and $\Tr\sigma=1$. For this state, the blocks in Eq.~\eqref{eq:choi-blocks} reduce to
\begin{equation*}
 C_0=aI_3,
 \qquad
 C_2=cI_3,
 \qquad
 C_1=
 \begin{pmatrix}
 b&\sqrt{ac}&\sqrt{ac}\\
 \sqrt{ac}&b&\sqrt{ac}\\
 \sqrt{ac}&\sqrt{ac}&b
 \end{pmatrix}.
\end{equation*}
The eigenvalues of $C_1$ are
\begin{equation*}
 b+2\sqrt{ac},
 \qquad
 b-\sqrt{ac},
 \qquad
 b-\sqrt{ac}.
\end{equation*}
Therefore the eigenvalues of $J(\Lambda_\sigma)^{T_B}$ are $a/3$ with multiplicity three, $c/3$ with multiplicity three, and
\begin{equation*}
 \frac{b+2\sqrt{ac}}{3},
 \qquad
 \frac{b-\sqrt{ac}}{3}
 \quad\text{with multiplicity two}.
\end{equation*}
They are all nonnegative whenever $b\geq\sqrt{ac}$. Thus $J(\Lambda_\sigma)$ is PPT. A PPT channel cannot generate distillable entanglement across the sender--receiver cut, even under arbitrarily many parallel uses; hence its unassisted quantum capacity vanishes \cite{Horodecki1998prl,Smith2012prl}.
\end{proof}

For example, $a=c=1/4$ and $b=1/2$ give
\begin{equation*}
 \sigma_*=\begin{pmatrix}
 1/4&0&1/4\\
 0&1/2&0\\
 1/4&0&1/4
 \end{pmatrix},
\end{equation*}
for which $b>\sqrt{ac}=1/4$. Hence $J(\Lambda_{\sigma_*})$ is PPT and $Q(\Lambda_{\sigma_*})=0$.

\begin{remark}\label{rem:ppt-private}
The PPT mechanism proves zero quantum capacity but does not by itself imply zero private capacity. This differs from the antidegradability mechanism in Corollary~\ref{cor:parity-invariant} and Proposition~\ref{prop:reflection-antidegradable}, which forces both capacities to vanish.
\end{remark}

\section{Positive coherent information and positive quantum capacity}\label{sec:positive}

The previous section identifies broad zero-capacity classes, but the reflection identity does not imply that every Hadamard-convolution channel is antidegradable. We now give a one-qutrit example with positive coherent information.

\begin{proposition}[Positive coherent information]\label{prop:positive}
Let $d=3$, $n=1$, and
\begin{equation}
 \ket\phi=\sqrt{\frac13}\ket0+\sqrt{\frac23}\ket1,
 \qquad
 \sigma=\proj\phi.
\label{eq:positive-environment}
\end{equation}
For the input state
\begin{equation}
 \rho_0
 =\frac34\ket0\bra0+\frac14\ket1\bra1
  +\frac14\ket0\bra1+\frac14\ket1\bra0,
\label{eq:positive-input}
\end{equation}
one has
\begin{equation}
 I_c(\rho_0,\Lambda_\sigma)
 :=S(\Lambda_\sigma(\rho_0))-S(\Lambda_\sigma^c(\rho_0))
 =0.046831089\ldots>0.
\label{eq:positive-Ic}
\end{equation}
Consequently, $Q(\Lambda_\sigma)>0$.
\end{proposition}

\begin{proof}
The state $\ket\phi$ has exactly two nonzero computational-basis amplitudes and is therefore nonstabilizer \cite{Gross2006jmp}. Its support is not invariant under parity, so it is also asymmetric. The nonzero principal block of $\rho_0$ is
\begin{equation*}
 \frac{1}{4}\begin{pmatrix}3&1\\1&1\end{pmatrix},
\end{equation*}
which has unit trace and determinant $1/8>0$; hence $\rho_0$ is a density operator.

Using Eqs.~\eqref{eq:direct-kraus} and \eqref{eq:complement-kraus}, a direct calculation gives
\begin{equation}
 \Lambda_\sigma(\rho_0)
 =\frac1{12}
 \begin{pmatrix}
 3&\sqrt2&0\\
 \sqrt2&2&0\\
 0&0&7
 \end{pmatrix},
 \qquad
 \Lambda_\sigma^c(\rho_0)
 =\frac1{12}
 \begin{pmatrix}
 5&0&0\\
 0&6&\sqrt2\\
 0&\sqrt2&1
 \end{pmatrix}.
\label{eq:positive-outputs}
\end{equation}
Their spectra are
\begin{align}
 \spec\Lambda_\sigma(\rho_0)
 &=\left(\frac1{12},\frac13,\frac7{12}\right),
 \label{eq:direct-spectrum}\\
 \spec\Lambda_\sigma^c(\rho_0)
 &=\left(\frac5{12},
          \frac{7+\sqrt{33}}{24},
          \frac{7-\sqrt{33}}{24}\right).
 \label{eq:complement-spectrum}
\end{align}
Writing $H(p_1,p_2,p_3)=-\sum_jp_j\log_2p_j$, we obtain the exact entropy expression
\begin{align*}
 I_c(\rho_0,\Lambda_\sigma)
 ={}&H\!\left(\frac1{12},\frac13,\frac7{12}\right)-H\!\left(\frac5{12},
          \frac{7+\sqrt{33}}{24},
          \frac{7-\sqrt{33}}{24}\right)={}0.046831089\ldots>0.
\end{align*}
It follows from Eqs.~\eqref{eq:coherent-one-shot} and \eqref{eq:regularized-capacities} that
\begin{equation*}
 Q(\Lambda_\sigma)
 \geq Q^{(1)}(\Lambda_\sigma)
 \geq I_c(\rho_0,\Lambda_\sigma)>0.
\end{equation*}
\end{proof}

Coherent information is continuous in both the input state and the channel in finite dimension. Positivity therefore persists in an open neighborhood of the pair $(\rho_0,\sigma)$. Proposition~\ref{prop:reflection-antidegradable} also implies that the environment in Proposition~\ref{prop:positive} cannot be channel-reflection equivalent. This example establishes that the parity-reflection identity is a conditional zero-capacity mechanism rather than a universal antidegradability statement.

\section{Conclusion and outlook}\label{sec:conclusion}

We have established an exact parity-reflected complementarity identity for finite-dimensional Hadamard-convolution channels. The weak complementary channel generated by an environment is the same convolutional channel generated by the parity-reflected environment; for pure environments, the statement applies to the ordinary complementary channel. This identity turns symmetries of the environmental state into explicit channel-level antidegradability criteria.

The identity gives rise to several robust zero-capacity mechanisms. If the environmental state is parity-invariant, then the corresponding channel is antidegradable, and if the environmental state is pure parity-invariant, the channel is self-complementary. More generally, channel-reflection-equivalent environments also give antidegradable channels. The asymmetric nonstabilizer state in Example~\ref{ex:magic-zero} belongs to this larger class. A separate family of qutrit PPT channels provides a second zero-quantum-capacity mechanism. At the same time, Proposition~\ref{prop:positive} gives an asymmetric nonstabilizer environment with positive coherent information and hence positive quantum capacity.

The central conclusion is that environmental magic and channel capacity are not related by a state-only criterion. The same Hadamard-convolution coupling admits asymmetric nonstabilizer environments that generate antidegradable zero-capacity channels and other asymmetric nonstabilizer environments that yield positive quantum capacity. What distinguishes these cases is the channel-level relation between the original and parity-reflected environments.

For a pure environment, Theorem~\ref{thm:reflection} reduces the complete antidegradability problem to the existence of a quantum channel $\mathcal D$ satisfying
\begin{equation}
 \Lambda_\sigma
 =\mathcal D\circ\Lambda_{\Pi\sigma\Pi^\dagger}.
\label{eq:open-antidegrading}
\end{equation}
Channel-reflection equivalence applies to the special case where $\mathcal D$ is unitary. Characterizing all pure states for which a general completely positive trace-preserving $\mathcal D$ exists remains open. In mixed environments, the corresponding classification must also retain the purification reference that appears in the ordinary complementary channel.

\backmatter

\section*{Statements and Declarations}

\subsection*{Funding}
L.L. was supported by the Research Startup Funding of Hangzhou City University under Grant No.~QD2024012. C.X. was supported by the National Natural Science Foundation of China under Grant No.~12201555 and the Natural Science Foundation of Hunan Province under Grant No.~2025JJ50050.

\subsection*{Competing interests}
The authors declare that they have no financial or non-financial interests that are directly or indirectly related to the work submitted for publication.

\subsection*{Author contributions}
Both authors contributed to the conception of the work, the mathematical analysis, and the preparation of the manuscript. Both authors read and approved the final manuscript.

\subsection*{Data availability}
No datasets were generated or analyzed during the current study.

\subsection*{AI Tool Use Declaration}

During manuscript preparation, the authors used OpenAI ChatGPT (GPT-5.6 Sol) to assist with language editing, checking and reformulation of some derivations. All AI-assisted mathematical content was independently verified and revised by the authors, who take full responsibility for the results and the final manuscript.


\begin{thebibliography}{99}



\bibitem{DiVincenzo1998pra}
DiVincenzo, D.P., Shor, P.W., Smolin, J.A.: Quantum-channel capacity of very noisy channels. Phys. Rev. A \textbf{57}, 830--839 (1998). \url{https://doi.org/10.1103/PhysRevA.57.830}

\bibitem{Leditzky2018prl}
Leditzky, F., Leung, D., Smith, G.: Dephrasure channel and superadditivity of coherent information. Phys. Rev. Lett. \textbf{121}, 160501 (2018). \url{https://doi.org/10.1103/PhysRevLett.121.160501}

\bibitem{Smith2008science}
Smith, G., Yard, J.: Quantum communication with zero-capacity channels. Science \textbf{321}, 1812--1815 (2008). \url{https://doi.org/10.1126/science.1162242}

\bibitem{Smith2012prl}
Smith, G., Smolin, J.A.: Detecting incapacity of a quantum channel. Phys. Rev. Lett. \textbf{108}, 230507 (2012). \url{https://doi.org/10.1103/PhysRevLett.108.230507}

\bibitem{Horodecki1998prl}
Horodecki, M., Horodecki, P., Horodecki, R.: Mixed-state entanglement and distillation: Is there a ``bound'' entanglement in nature? Phys. Rev. Lett. \textbf{80}, 5239--5242 (1998). \url{https://doi.org/10.1103/PhysRevLett.80.5239}

\bibitem{Bu2023pnas}
Bu, K., Gu, W., Jaffe, A.: Quantum entropy and central limit theorem. Proc. Natl. Acad. Sci. USA \textbf{120}, e2304589120 (2023). \url{https://doi.org/10.1073/pnas.2304589120}

\bibitem{Bu2023arxiv}
Bu, K., Gu, W., Jaffe, A.: Discrete quantum Gaussians and central limit theorem. arXiv:2302.08423 (2023). \url{https://doi.org/10.48550/arXiv.2302.08423}

\bibitem{Bu2025prl}
Bu, K., Jaffe, A.: Magic resource can enhance the quantum capacity of channels. Phys. Rev. Lett. \textbf{134}, 050202 (2025). \url{https://doi.org/10.1103/PhysRevLett.134.050202}

\bibitem{Sun2025pra}
Sun, W., Jin, Y.: Leveraging magic resources for quantum channel capacity enhancement under stabilizer convolution. Phys. Rev. A \textbf{112}, 052412 (2025). \url{https://doi.org/10.1103/qnz3-4fsc}

\bibitem{Xiong2026arxiv}
Xiong, C., Kim, S., Long, L., Wu, J.: Private capacity of quantum channels induced by non-stabilizer environmental states. arXiv:2607.11793 (2026). \url{https://doi.org/10.48550/arXiv.2607.11793}

\bibitem{Lloyd1997pra}
Lloyd, S.: Capacity of the noisy quantum channel. Phys. Rev. A \textbf{55}, 1613--1622 (1997). \url{https://doi.org/10.1103/PhysRevA.55.1613}

\bibitem{Shor2002msri}
Shor, P.W.: The quantum channel capacity and coherent information. Lecture notes, MSRI Workshop on Quantum Computation, Berkeley (2002).

\bibitem{Devetak2005tit}
Devetak, I.: The private classical capacity and quantum capacity of a quantum channel. IEEE Trans. Inf. Theory \textbf{51}, 44--55 (2005). \url{https://doi.org/10.1109/TIT.2004.839515}

\bibitem{Cubitt2008jmp}
Cubitt, T.S., Ruskai, M.B., Smith, G.: The structure of degradable quantum channels. J. Math. Phys. \textbf{49}, 102104 (2008). \url{https://doi.org/10.1063/1.2953685}

\bibitem{Bradler2010pra}
Br{\'a}dler, K., Hayden, P., Touchette, D., Wilde, M.M.: Trade-off capacities of the quantum Hadamard channels. Phys. Rev. A \textbf{81}, 062312 (2010). \url{https://doi.org/10.1103/PhysRevA.81.062312}

\bibitem{Wilde2014cmp}
Wilde, M.M., Winter, A., Yang, D.: Strong converse for the classical capacity of entanglement-breaking and Hadamard channels via a sandwiched R{\'e}nyi relative entropy. Commun. Math. Phys. \textbf{331}, 593--622 (2014). \url{https://doi.org/10.1007/s00220-014-2122-x}

\bibitem{Gross2006jmp}
Gross, D.: Hudson's theorem for finite-dimensional quantum systems. J. Math. Phys. \textbf{47}, 122107 (2006). \url{https://doi.org/10.1063/1.2393152}

\end{thebibliography}
\end{document}